\documentclass[12pt]{iopart}

\usepackage{graphicx}

\begin{document}

\title{Magnon Bose condensation in symmetry breaking magnetic field}

\author{S.V. Maleyev, V.P. Plakhty, S.V. Grigoriev, A.I. Okorokov and  A.V. Syromyatnikov}

\address{ Petersburg Nuclear Physics Institute, Gatchina, Leningrad District 188300, Russia }
\ead{maleyev@SM8283.edu.spb}
\begin{abstract}
Magnon Bose condensation (BC) in the symmetry breaking magnetic field is a result of unusual form of the Zeeman energy that has terms linear in the spin-wave operators and terms mixing excitations which momenta differ in the wave-vector of the magnetic structure. The following examples are considered: simple easy-plane tetragonal antiferromagnets (AFs), frustrated AF family R$_2$CuO$_4$, where R=Pr, Nd etc., and cubic magnets with the Dzyaloshinskii-Moriya interaction (MnSi etc.). In all cases the BC is important when the magnetic field is comparable with the spin-wave gap. The theory is illustrated by existing experimental results.  
\end{abstract}

%Uncomment for PACS numbers title message
%\pacs{00.00, 20.00, 42.10}
% Keywords required only for MST, PB, PMB, PM, JOA, JOB? 
%\vspace{2pc}
%\noindent{\it Keywords}: Article preparation, IOP journals
% Uncomment for Submitted to journal title message
%\submitto{\JPA}
% Comment out if separate title page not required
%\maketitle

\section{Introduction}

Magnon Bose condensation (BC) in magnetic field was intensively studied in spin singlet materials (see for example [1] and references therein). In this case magnons condens in the field just above the triplet gap. In this paper we consider magnon BC that appears in the symmetry breaking magnetic field. The theoretical discussion is illustrated by experimental observation of this BC in frustrated antiferromagnet (AF) Pr$_2$CuO$_4$ and cubic helimagnets MnSi and FeGe. To clarify our idea we begin with  consideration of conventional AFs.  In textbooks  two limiting cases are considered. First, the magnetic field is directed along the sublattices. In this case the system remains stable up to the critical field $H_C=\Delta$, where $\Delta$ is the spin-wave gap. Then the first order transition occurs to the state in which the field is perpendicular to sublattices (spin-flop transition). Second, the field is perpendicular to initial staggered magnetization. The system remains stable but the spins are canted toward the field by the angle determined by $\sin\vartheta=-H/(2S J_0)$, where $J_0=J z$; $J$ and $z$ are the exchange interaction and the number of nearest neighbors, respectively. At $H+2S J_0$ the spin-flip transition occurs to the ferromagnetic state. To the best of our knowledge the first consideration of the symmetry breaking field was performed theoretically in [2] in connection with experimental study of the magnetic structure of the frustrated AF R$_2$CuO$_4$, where R=Pr, Nd, Sm and Eu [3,4]. In these papers the non-collinear structure was observed using the neutron scattering in the field directed at angle of $\delta=45^0$ to the sublattices. It was found in [2] that in inclined field the Zeeman energy has unusual form with terms which are linear in the spin-wave operators and term mixing magnons which momenta differ in the AF vector $\bi k_0$. As a result the BC arises of the spin-waves with momenta equal to zero and $\pm\bi k_0$.
 
 Similar situation exists in cubic helimagnets  MnSi etc. [5].  If the field is directed along the helix wave-vector $\bi k$ the plain helix transforms into conical structure and then the ferromagnetic spin state occurs at critical field $H_C$. But if $\bi H \perp \bi k$ the magnon condense with momenta zero, $\pm\bi k$, $\pm 2\bi k$ etc. This leads to the following observable phenomena: i)  a transition to the state  with $\bi k$ directed along the field at $H_\perp\sim H_{C1}=\Delta\sqrt2$, where $\Delta $ is the spin-wave gap, ii) the second harmonic  of the spin rotation with the vector $2\bi k$ and perpendicular spin susceptibility at $H_\perp<H_{C1}$. Rotation of the helix was observed in [6-8].

In this paper we outline  basic ideas of the BC in the symmetry breaking field as applied to the frustrated cuprates and  non-centrosymmetric cubic helimagnets. We illustrate  the theory with some  recent experiments.

\section{Non-frustrated AF}

To demonstrate basic ideas of our approach we begin with non-frustrated easy-plane tetragonal AF. We are not interested in thermal fluctuations and consider  single AF plane. If the field is directed at angle $\delta$ to the $b$ axis (see figure 1)  sublattices rotate by the angle $\varphi$. Simultaneously, the field component perpendicular to new $Z$ axis cants the spins by the angle $\vartheta\simeq-H_\perp/(2S J_0)\ll1$. As a result we have for two neighboring spins in $(ZY)$ frame [2]
\begin{equation}
	\bi S_1=S_{\rm z1}\hat Z+\vartheta S_{\rm y1}\hat Y;\quad \bi S_2=-S_{\rm z2}\hat Z+\vartheta S_{\rm y 2}\hat Y, \end{equation}
	where in the linear spin-wave theory $S_{\rm z l}=S-a^+_la_l$ and $S_{\rm y l}=-\rmi \sqrt{S/2}(a_l-a^+_l)$,  $l=1,2$ and $a_l(a^+_l)$ are Bose operators. As a result the Zeeman energy has unusual form
\begin{equation}
	H_Z=H_\parallel\sum a^+_{\bi q+\bi k_0}a_\bi q+\rmi \vartheta\sqrt{N S/2} (a_0-a^+_0),
\end{equation}
where $\bi k_0$ and $N$ are the AF wave-vector and total spin number, respectively. Here the first term mixes spin-waves with momenta $\bi q$ and $\bi q\pm\bi k_0$ and the second one excites (absorbs) magnons with $\bi q=0$.
Along with this energy we have conventional spin-wave Hamiltonian $H_{\rm{SW}}=\sum[E_{\bi q}a^+_\bi q a_\bi q+B_\bi q(a_{-\bi q}a_\bi q+a^+_{-\bi q}a^+_\bi q)/2]$ with the spin-wave energy $\epsilon_\bi q=(E^2_\bi q-B^2_\bi q)^{1/2}$.
The spin-wave gap is $\epsilon_0=\Delta$.

Linear terms in Eq.(2) contribute to the ground state energy if $a_0(a^+_0)\sim \sqrt N$, i.e.  these operators  has to be considered as classical variables as in the Bogoliubov theory of the BC in dilute Bose gas. Due to the first term in (2) we must consider  the operators $a_{\pm\bi k_0}$ and  $a^+_{\mp\bi k_0}$ as classical variables too. Minimizing the full Hamiltonian with respect to these variables we obtain
\begin{equation}
	E=(\Delta^2\sin^22\varphi)/(16J_0)-S^2J_0 \vartheta^2-(H_\parallel H_\perp)^2/[4J_0(\Delta^2(\varphi,\bi H)],
\end{equation}
where the first term is the energy of the square anisotropy. In cuprates with $S=1/2$ it has quantum origin and  arises due to pseudodipolar in-plane interaction [9]. The second term is the energy of the spin canting in perpendicular field. The  last term is the BC energy and 
$\Delta^2(\varphi,\bi H) = \Delta^2\cos4\varphi+H^2_\perp-H^2_\parallel$ 
is the spin-wave gap in the  field [2]. This contribution becomes important at $H\sim \Delta$. The spin configuration is determined by $\rmd E/\rmd\varphi=0$ and equilibrium condition $\rmd^2 E/\rmd \varphi^2\geq 0$.

This theory was verified by neutrons scattering [10,11]. In diagonal field $H\parallel(1,1,0)$ the spin configuration in frustrated Pr$_2$CuO$_4$ is governed by Eq.(3) and the intensity of the $(1/2,1/2,-1)$ is given by $I\sim1+\sin2\varphi$ [2]. Neglecting the BC term we get $\sin2\varphi=-(H/H_C)^2$, where  $H_C=\Delta$. As a result at $H\to H_C$ we obtain $I\sim H_C-H$. But very close to $H_C$ the BC term becomes important and we have a crossover to $I\sim(H_C-H)^{1/2}$. It is clearly seen in figure 2. This crossover was observed in [10,11].

\section{Frustrated AFs}

In frustrated R$_2$CuO$_4$ AFs there are two copper spins in unit cell belonging to different CuO$_2$ planes (see inset in figure 1). From symmetry considerations these spins do not interact in the exchange approximation. The orthogonal spin structure is a result of the interplane pseudodipolar interaction (PDI) [2,3] and the ground state energy is given by
\begin{eqnarray}
\fl	E=\frac{\Delta^2}{16J_0}[\sin^2 2\varphi_1+\sin^22\varphi_2-4G\sin(\varphi_1+\varphi_2)]-S^2J_0(\vartheta^2_1+\vartheta^2_2)+E_C(\varphi_1,\varphi_2,\bi H),
\end{eqnarray}
where $\vartheta_{12}=-H_{\perp12}/(2S J_0)$, $G=(\Omega/\Delta)^2$ and $\Omega^2$ is a difference between square of optical and acoustic spin-wave branches at $H=0$. The BC energy $E_C$ has very complicated form  [2] and we do not present it here.

For Pr$_2$CuO$_4$ at $T=18K$ we have $\Delta\simeq 0.36meV$, $\Omega\simeq 2.8meV$ and $\Omega\simeq 60\ll 1$ [2]. Then the intraplane PDI is strong and the BC contribution can not be neglected at low field $H<\Delta$. We illustrate this by results of particular calculations taking and neglecting BC in the field almost along the $b$ axis ($\delta\ll1$). Instead of  $\varphi_{1,2}$ we  use new angles determined as $\varphi_1=\alpha+\gamma/2,\quad \varphi_2=-\pi/2-\alpha+\gamma/2$. Neglecting the BC we have $\alpha=-(H/\Delta)^2\delta$ and $\gamma=(H/\Delta)^4\delta/G$. The BC changes the last result: $\gamma_{\rm B\rm C}=\gamma G\gg \gamma$.

 Role of the BC can be illustrated by results of neutron scattering in Pr$_2$CuO$_4$ [12]. The angles $\alpha$ and $\gamma$ were determined from measurements of two  reflections $(1/2,1/2,1)$ and $(-12,1/2,1)$. If $\delta=0$
 the zero field spin configuration remains stable at $H<H_C\Delta G^{1/4}$ and there is no BC as $H_{\parallel1}=H_{\perp2}=0$ [see Eq.(2)]. Then the theory without BC predicts the firs order transition to the collinear non-spin-flop  state with $\alpha=_45^0$ and $\tan \gamma_C=G^{1/2} $. This transition is seen in figure 3 at $H_C\simeq 6.5T$ [12]. From these data we obtain $\gamma_C\simeq 30^0$. Using parameters given above and neglecting the BC we  obtain $H_C\simeq 6.7T$ and $\gamma_C\simeq 7.4^0$. The last quantity is in strong disagreement with experiment. It was demonstrated in [13] that $\Delta$ depends on temperature  and at $T=10K$ we have $\Delta\simeq 0.5meV$ as in figure 3. Assuming that $\Omega$ does not depend on $T$ we obtain $H_C\simeq 7.8T$ and $\gamma_C\simeq 5.3^0$ that is in stronger disagreement wit the experiment.
 
 The experimentally obtained angles $\alpha$ and $\gamma$  at $T=18K$ and $\delta=9.5^0$ are shown in figure 4 [12]. The transition to the collinear state with $\alpha\sim -45^0$ and $\gamma_C\sim 20^0$ was observed. Again the non-BC theory can not explain the  experimental data. For example it gives $\gamma_C\simeq 2.5^0$. Explanation of all these experimental data using the BC theory will be  given elsewhere.
 
 \section{BC in helimagnets}
 
In helimagnets MnSi etc. Dzyaloshinskii-Moriya interaction (DMI) stabilizes the helical structure and the  helix wave-vector  has the form $\bi k=S D[\hat a\times \hat b]/A$, where $D$ is the strength of the DMI, $A$ is the spin-wave stiffness at momenta $q\gg k$, $\hat a$ and $\hat b$ are unit orthogonal vectors in the plane of the spin rotation.
 
 The classical energy depends on the field component $H_\parallel$ along the vector $\bi k$ and the cone angle of the spin rotation is given by $\sin\alpha=-H/H_C$, where $H_C=A k^2$ is the critical field of the transition into ferromagnetic state [5].  However at $H_\perp\ll H_C$ rotation of the helix axis toward the field direction and the second harmonic $2\bi k$ of the spin rotation were observed [6-8].  Both phenomena are related to the magnon BC in perpendicular field [5].
 
The linear and mixing terms appear in the Zeeman energy in much the same way as it was discussed above:
\begin{equation}
\fl	H_Z=(H_a-\rmi H_b)\sqrt{NS/2}(a_{-\bi k}-a^+_\bi k)/2
-\sum (a^+_\bi k a_0+a^+_0 a_{-\bi k})+H.C.,
\end{equation}
and we have the magnon BC with momenta zero and $\pm\bi k$. Corresponding contribution to the ground state energy is given by
\begin{equation}
	E_C=-SH^2_\perp \Delta^2/[H_C(\Delta^2-H^2_\perp/2)].
\end{equation}
  Obviously near the critical point $H_\perp=\Delta\sqrt 2$ the real form of the BC energy is not so simple. It is determined by nonlinear interactions but consideration of this problem is out of the scope of this paper.
  
  As a result the perpendicular susceptibility is proportional to $1/(\Delta^2-H^2_\perp/2)$ and $2\bi k$ harmonic appears. The last was observed by neutron scattering [6-8]. Intensities  of corresponding Bragg satellites  have the form
\begin{equation}
	I_\pm \sim[\Delta^2/(\Delta^2-H^2_\perp/2)]^2[1\mp(\bi k \bi P)]\delta(\bi q\mp 2\bi k),
\end{equation}
  where $\bi P$ is the neutron polarization.

  If $H_\perp\to \Delta\sqrt 2$ the helix axis rotates toward the field. This rotation is governed by competition of the BC and crystallographic energies [5]. Evolution of the Bragg reflections in MnSi with $H_\perp$ is shown in figure 5. 

\section{Conclusions}

  We discuss a few examples of the  magnon BC in symmetry breaking  magnetic field. BC appears due to unusual terms in Zeeman energy. Obviously this phenomenon is very general  and can be observed in other ordered magnetic systems. Effects related to the BC  has to be more pronounced in the field of order of the sin-wave gap. 
  
  \section{Acknowledgments} 
  
  This work is partly supported by RFBR (Grants 03-02-17340, 06-02-16702 and 00-15-96814), Russian state programs "Quantum Macrophysics", "Strongly Correlated electrons in Semiconductors, Metals, Superconductors and  Magnetic Materials" "Neutron Research of Solids", Japan-Russian collaboration 05-02-19889-JpPhysics-RFBR and Russian Science Support Foundation (A.V.S.).

 \section*{Figures} 
 
\begin{figure}
\centering
\includegraphics[scale=0.7]{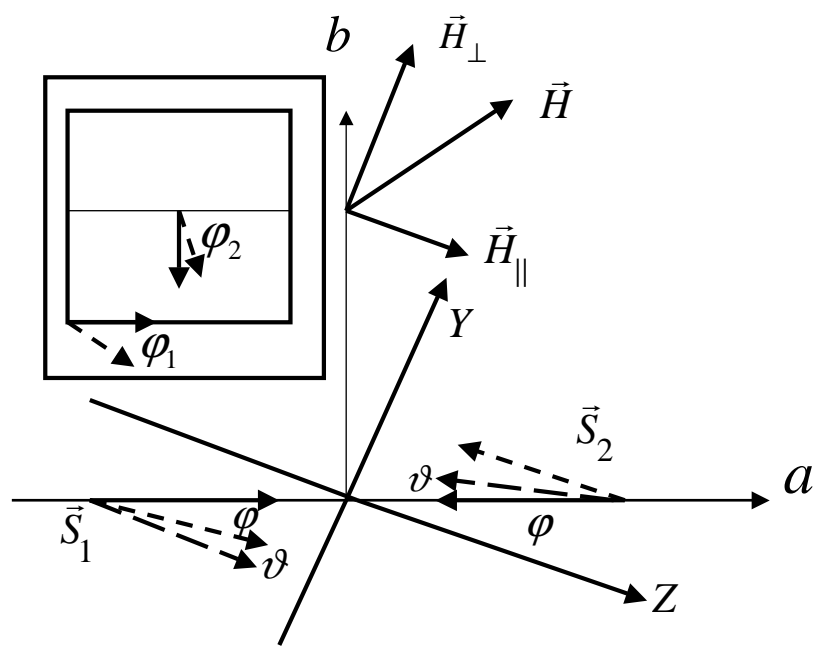}
\caption{Spin configuration in the field. Full and dashed arrows correspond to zero and nonzero field, respectively. Addition  spin canting in $H_\perp$ is shown by broken arrows. Inset: spin configuration in neighboring planes of frustrated AF.} 
\end{figure}

\begin{figure}
\centering
\includegraphics[scale=0.5]{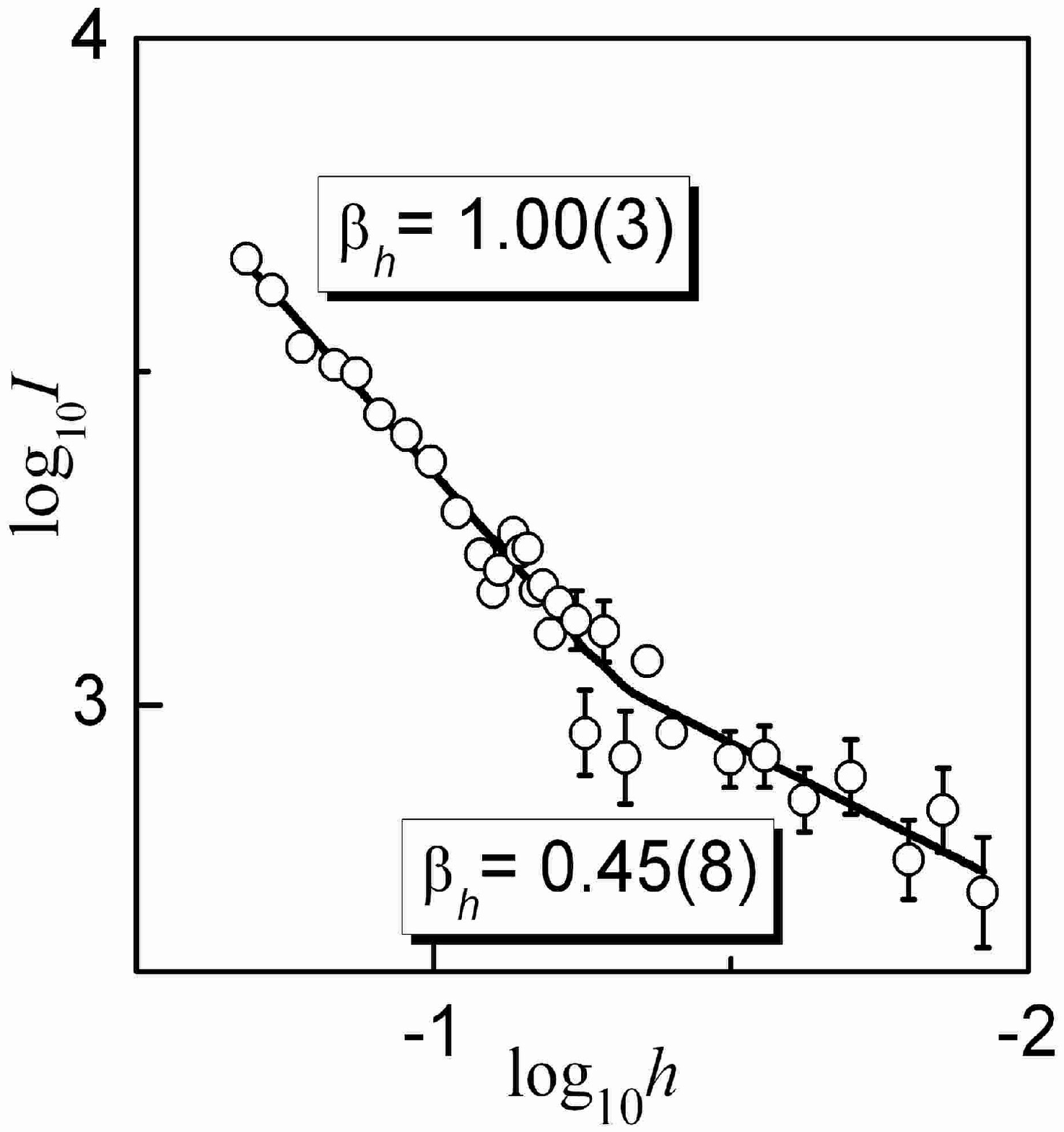}
\caption{Log-Log plot of the $(1/2,1/2,-1)$ Bragg intensity in diagonal field, $h\sim (H_C-H)$.} 
\end{figure}

\begin{figure}
\centering
\includegraphics{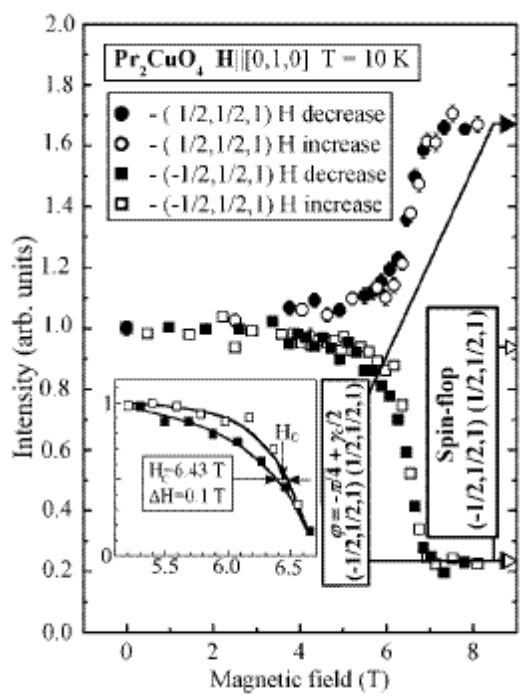}
\caption{The first order transition in the field directed along $b$ axis. Calculated intensities for the spin flop configurations when spins are perpendicular to the field (white arrows).} 
\end{figure}

\begin{figure}
\centering
\includegraphics[scale=0.7]{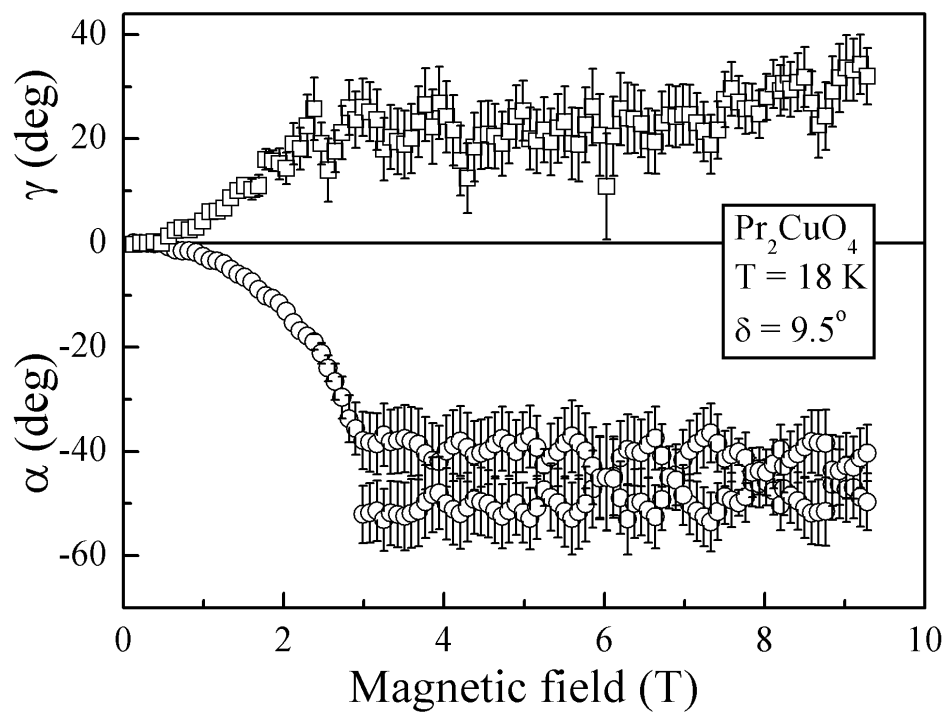}
\caption{Field dependence of angles $\alpha$ and $\gamma$  at $\delta=9.5^0$.} 
\end{figure}

\begin{figure}
\centering
\includegraphics[scale=0.7]{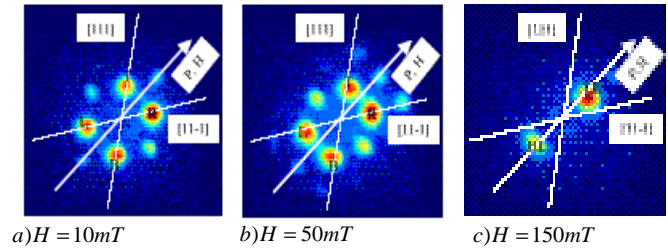}
\caption{Bragg reflections in the field along $(1,1,0)$. a) Four strong spots corresponds to $\pm (1,1,1)$ and $\pm(1,1,-1)$ reflections. Weak spots are the double Bragg scattering. b) The $2\bi k$ satellites appear. c) The helix vector is directed along the field.} 
\end{figure}

\end{document}